\documentclass[12pt]{article}
\title{ Analytic calculation of field-strength correlators}
\author{Yu.A.Simonov\\
 State Research
Center\\Institute of Theoretical and Experimental Physics, \\
Moscow, 117218 Russia}
 \date{}
\newcommand{\beq}{\begin{eqnarray}}
 \newcommand{\eeq}{\end{eqnarray}}
\newcommand{\be}{\begin{equation}}
 \newcommand{\ee}{\end{equation}}

 \def\la{\mathrel{\mathpalette\fun <}}

\def\fun#1#2{\lower3.6pt\vbox{\baselineskip0pt\lineskip.9pt
\ialign{$\mathsurround=0pt#1\hfil ##\hfil$\crcr#2\crcr\sim\crcr}}}

\newcommand{{\SD}}{\rm SD}

\newcommand{\ver}{\mbox{\boldmath${\rm r}$}}

\newcommand{\vep}{\mbox{\boldmath${\rm p}$}}

\newcommand{\vez}{\mbox{\boldmath${\rm z}$}}
\newcommand{\veS}{\mbox{\boldmath${\rm S}$}}

\newcommand{\vexi}{\mbox{\boldmath${\rm \xi}$}}
\newcommand{\veta}{\mbox{\boldmath${\rm \eta}$}}

\newcommand{\llan}{\langle\langle}
\newcommand{\rran}{\rangle\rangle}
\newcommand{\lan}{\langle}
\newcommand{\ran}{\rangle}
\begin{document}
\maketitle

\begin{abstract}

 Field correlators are expressed using background field formalism through
the gluelump Green's functions. The latter are obtained in the
path integral and Hamiltonian formalism.  As a result behaviour of
  field correlators  is obtained  at small and
large distances both for perturbative and nonperturbative parts.
The latter decay exponentially at large distances and are finite
at $x=0$, in agreement with OPE and lattice data.

 \end{abstract}

\section{Introduction}

The Method of Field Correlators (MFC), suggested some years ago
\cite{1} (see \cite{2} for a review) has successfully produced a
large number of results in nonperturbative QCD, in particular
predicting hadron masses in good agreement  with  lattice and
experimental data (see e.g. \cite{3}  for a review) and
unambigiously  explaining linear confinement \cite{4}, \cite{5}.

In doing so MFC is exploiting  the field  correlators known from
lattice data \cite{6} as input parameters. Recent discovery of
Casimir scaling  for static   potentials \cite{7,8} allows to
neglect all correlators except for the lowest ones (bilocal or
Gaussian) with accuracy about 1\% and this fact reduces input
parameters essentially to the string tension $\sigma$ for the bulk
of the hadron spectrum and in addition correlation length $T_g$
 for hadron spin
splittings. Thus all spin-averaged spectrum of hadrons, namely
light mesons \cite{9}, heavy quarkonia \cite{10}, hybrids
\cite{3,11,12}, heavy-light mesons \cite{13}, glueballs \cite{14}
and baryons \cite{15} are calculated through the only parameter
$\sigma$ in good agreement $(\sim 10\%)$ with experimental and
lattice data.

Similar situation occurs for the background perturbation theory in
the real QCD vacuum, where the only additional parameter $m_B$,
calculated in terms of $\sigma$, can be used to construct a new
improved perturbation series without Landau ghost pole and  IR
renormalons \cite{16}. With all that the situation in MFC is not
yet satisfactory, since till now field correlators have not been
computed analytically within the method itself. The first attempts
in this direction have been done in \cite{17} where equations for
field correlators have been first written and the correlation
length $T_g$ was computed in terms of $\sigma$, in reasonable
agreement with lattice data \cite{6}. (For earlier developments in
the framework of stochastic quantization method see \cite{18}.
Resulting equations there however contain path integrals and are
too complicated for practical use).

Recently new objects - gluelumps - have been introduced \cite{19},
which represent bound states of a valence gluon in the field of
static gluonic charge -- an adjoint equivalent of heavy-light
mesons. Spectrum of gluelumps was calculated on the lattice
\cite{20} and analytically in \cite{21}, being in general
agreement with each other.

It  was realized subsequently, that gluelump Green's functions and
field correlators are the same, if in - and out - states of
gluelumps are constructed with the help of field strength
operators.

For the analytic calculations in \cite{21} the Background Field
Formalism (BFF)  was used \cite{22,16}, which allows to construct
gluelump states and the Hamiltonian in a very simple local form.
Essentially the valence gluon operators have been introduced in
\cite{21} and used to construct the whole hierarchy of gluelump
states, which agrees resultatively with the states used in lattice
calculations \cite{20}. Therefore one can use  now the simple
technic of \cite{21} to calculate the gluelump Green's functions,
and through them, the field correlators. In this way the logic of
the method is closed, since in principle one can calculate all
field correlators and through them all physical quantities in
terms of only one input parameter - the string tension. In
practice calculations are approximate and use Gaussian
approximation, when only lowest correlators are retained with
later improvements.

The full implementation of this program requires solution of three
basic problems:
\begin{enumerate}
\item  Dynamical definitions of  perturbative and nonperturbative.
In terms of BFF one needs to understand how the nonperturbative
fields are constructed in contrast to the  perturbative valence
gluon fields.

\item Gaussian dominance, i.e. suppression of contributions of
quartic and higher field correlators.

\item Explicit relation between $\Lambda_{QCD} $ and $\sigma$, so
that all resulting quantities containing both perturbative and
nonperturbative contributions, can be expressed through only one
scale parameter.
\end{enumerate}

In the present paper only  partial answers will be given to  the
questions 1 and  3, nevertheless allowing to pave the road for
future developments.

The paper is organized as follows. In section 2 the field
correlators are written in the framework of BFF and the correlator
$D_1$ is   expressed in terms of gluelump Green's functions. In
section 3 a similar  detailed study of the confining correlators
$D(x)$ is done and its behaviour is defined at small and large
$x$. In section 4 properties and selfconsistency of field
correlators are discussed. Five appendices contain details of
derivation of basic equations in the main text.

\section{Field correlators in the Background Formalism}

We start with the standard definitions of the field correlator
functions $D(x)$ and $D_1(x)$, defined as  in \cite{1} $$
\frac{g^2}{N_c}\lan tr_f (F_{\mu\nu} (x) \Phi (x,y)
F_{\lambda\sigma} (y) \Phi (y,x))\ran \equiv
D_{\mu\nu,\lambda\sigma} (x,y)= $$ \be=
(\delta_{\mu\lambda}\delta_{\nu\sigma}-\delta_{\mu\sigma}\delta_{\nu\lambda})
D(x-y) +\frac12\left( \frac{\partial}{\partial x_\mu}
h_\lambda\delta_{\nu\sigma} +perm.\right) D_1(x-y)\label{1}\ee
where $h_\mu=x_\mu-y_\mu,~~\Phi(x,y)=P\exp ig \int^x_y A_\mu d
z_\mu$ and $tr_f$ is the trace in the fundamental representation.
Our final aim in this section will be to connect $D(x), D_1(x) $
to the gluelump Green's functions. If the latter are defined as it
is done on the lattice \cite{19,20}, i.e. with in- and out- states
constructed with the help of field strength operators, then the
connection is trivial
\be
D_{\mu\nu,\lambda\sigma} (x,y) = \frac{g^2}{2N_c^2} \lan tr_a\hat
F_{\mu\nu}(x) \hat \Phi(x,y) \hat F_{\lambda\sigma}
(y)\ran\label{2}\ee and here all dashed letters stand for
operators in the adjoint representation. These Green's functions
are however not accessible for analytic calculation and to proceed
one needs to use BFF \cite{16}, where the notions of valence gluon
field $a_\mu$ and background field $B_\mu$ are introduced, so that
total gluonic field $A_\mu$
 is written as
\be
A_\mu=B_\mu+a_\mu\label{3} \ee
 As it can be shown with the help of the 'tHooft identity \cite{16},
   the independent functional integration over $DB~Da$ does not
 contain the double counting, and one can proceed to define
 perturbation series in powers of ($ga_\mu)$ as it was done in
 \cite{16}. However for our purposes here we need a more explicit
 definition of the background and of field separation in
 (\ref{3}). To this end one can use the main idea suggested in
 \cite{17} which we explain here in the most simple form. Namely,
 let us single out some color index $a$ and  fix it at a given
 number. The field $A_\mu^a$ will be identified with $a^a_\mu$
 while the rest of fields will be called $A_\mu^b= B_\mu^b,~~
 b\neq a$. In the averaging over fields $B^b_\mu$, and only after
 this done one integrates also fields $a^a_\mu$ The essential
 point is as was shown in \cite{17} that the integration over
 $DB^b_\mu$ will provide a white adjoint string for the gluon
 $a^a_\mu$, which keeps the color index a unchanged. This is the
 basic physical mechanism  behind this background technic and it is
 connected to the properties of ensemble gluons: 1) even for
 $N_c=3$ one has one field $a^a_\mu$ and 7 fields $B^b_\mu$ 2)
 confining string is a colorless object and therefore the singled
 out color index $a$ can be preserved during interaction process
 of the valence gluon $a^a_\mu$ with the rest of gluons
 ($B^b_\mu)$.
These remarks make explicit the notions of the valence gluon and
background field and will be used in what follows.

Using (\ref{3}) one can write the total field operator
$F_{\mu\nu}(x)$ as follows $$ F_{\mu\nu} (x) =\partial_\mu
A_\nu-\partial_\nu A_\mu-ig [A_\mu,A_\nu]=$$ \be=
\partial_\mu(a_\nu+B_\nu)-\partial_\nu(a_\mu+B_\mu)-ig[a_\mu+B_\mu,a_\nu+B_\nu]=\label{4}
\ee $$=\hat D_\mu a_\nu-\hat D_\nu a_\mu-ig[a_\mu,a_\nu]+
F^{(B)}_{\mu\nu}.$$ Here the term, $F_{\mu\nu}^{(B)}$ contains
only the field $B_\mu^b$. It is clear that when one averages over
field $a_\mu^a $ and sums finally over all color indices  $a$, one
actually exploits all the fields with color indices from
$F^{(B)}_{\mu\nu}$, so that the term $F_{\mu\nu}^{(B)}$ can be
omitted, if summing over all $a$ is presumed to be done at the end
of calculation. In this section we shall concentrate on the first
two terms on the r.h.s. of (\ref{4}), leaving discussion of the
term $[a_\mu, a_\nu]$ to the section 3.

Assuming the background Feynman gauge, $ D_\mu a_\mu= 0,$
\cite{21} we shall define now the gluelump Green's function as
\be
G_{\mu\nu} (x,y) =\lan tr_a a_\mu(x) \hat \Phi(x,y)
a_\nu(y)\ran\label{5}\ee and the function
$D_{\mu\nu,\lambda\sigma}$ can be written as
\be
D_{\mu\nu,\lambda\sigma}(x,y) = D_{\mu\nu,\lambda\sigma}^{(0)}+
D_{\mu\nu,
\lambda\sigma}^{(1)}+D_{\mu\nu,\lambda\sigma}^{(2)}\label{6}\ee
where the superscript 0,1,2
 denotes the power of $g$ coming from the term $
(ig)[a_\mu,a_\nu]$ in (\ref{4}).

For the following one can use the relation $$ \lan tr_a (D_\mu
a_\nu(x)) \hat \Phi\chi\ran =\frac{\partial}{\partial x_\mu}\lan
tr_a a_\nu\hat \Phi\chi\ran-$$ \be-\lan tr_a a_\nu(x)
\delta_\mu(x) \hat \Phi \chi\ran \label{7} \ee where the following
notation is used for the  contour differentiation (see \cite{23}
for details and earlier refs.)
\be
\delta_\mu(x) \hat \Phi (x,y) =ig \int^x_y dz_\lambda \alpha(z,y)
\hat \Phi(x,z) \hat F_{\mu\lambda}(z) \hat \Phi(z,y).\label{8}\ee
Analogously for differentiation in the end point one has
\be
\hat \Phi(x,y) \overleftarrow{\delta_{\mu}}(y) =-ig \int^x_y
dz_\lambda \alpha(z,x) \hat \Phi (x,z) \hat F_{\mu\lambda} (z)
\hat \Phi(z,y)\label{9} \ee where $\alpha(z,y) =\left |
\frac{z-y}{x-y}\right|,~~ \alpha(z,x) =\left|
\frac{x-z}{x-y}\right|.$

One has $$ D_{\mu\nu,\lambda\sigma}^{(0)}(x,y) =
\frac{g^2}{2N^2_c}\left\{ \frac{\partial}{\partial
x_\mu}\frac{\partial}{\partial y_\lambda} \lan tr_a a_\nu (x) \hat
\Phi (x,y) a_\sigma(y)\ran+ \right.$$ $$+perm. -
\frac{\partial}{\partial y_\lambda} \lan tr_a a_\nu (x) \delta_\mu
(x) \hat \Phi a_\sigma(y)\ran -perm.$$ $$-\frac{\partial}{\partial
x_\mu} \lan tr_a a_\nu (x) \hat \Phi
(x,y)\overleftarrow{\delta_\lambda}(y) a_\sigma(y)\ran+$$
\be\left. + \lan tr_a a_\nu (x)\delta_\mu (x) \hat \Phi
(x,y)\overleftarrow{\delta_\lambda}(y) a_\sigma(y)\ran+ perm
\right\}.\label{10}\ee

From the structure of  the r.h.s. of   (\ref{10}) it is clear that
all terms except the last one (and its permutations) contribute to
the function $D_1$, which has the form of the full derivative, cf.
Eq.(\ref{1}). In what follows it is convenient to take  spacial
indices for $\mu,\nu$ in (\ref{5}), $\mu\to i, \nu\to k$, and
consider the correlator of color-electric fields $D_{4i, 4k}
(x,y)$, where $x,y$ are taken on the time axis. Hence the integral
in (\ref{8}), (\ref{9}) is over $dz_4$ and for $\mu=4$ both terms
(\ref{8}), (\ref{9}) disappear. We ~~ can~~ also ~~ write~~  for~~
the~~ gluelump ~~Green's~~~ function ~~~ (5) \\$G_{\mu\nu} (x,y) =
\delta_{\mu\nu} f ((x-y)^2)$.

As a result one obtains from (\ref{10}) the following connection
of $D^{(0)}$ and $f((x-y)^2)$
\be
 D^{(0)}_{4i,4k} (x,y) = \frac{g^2}{2N_c^2} \left \{\frac{\partial}{\partial
 x_4} \frac{\partial}{\partial y_4} \delta_{ik} f((x-y)^2)
  + \frac{\partial}{\partial x_i} \frac{\partial}{\partial y_k} f((x-y)^2)\right\}
\label{11}\ee
 on the other hand using  (1) with  $h_\mu\equiv x_\mu-y_\mu$ one  can express
 $D^{(0)}$ through $D_1$ as
 \be
 D_{4i,4k} (h) = \delta_{ik} D(h) +\frac12 \left(
 \frac{\partial}{\partial x_4} h_4 D_1 \delta_{ik} +
 \frac{\partial}{\partial x_i} h_k D_1\right) \label{12}\ee
 and for $h_i=0, i=1,2,3, h_4\neq 0$ one obtains
 \be
 D_1 (x) =- \frac{2g^2}{N^2_c}\frac{df(x^2)}{dx^2}
\label{13}\ee

To obtain information about the gluelump Green's function
$G_{\mu\nu}$ one can use the   path-integral representation of
$G_{\mu\nu}(x,y)$ in the Fock-Feynman-Schwinger (FFS) formalism
(see \cite{24} for reviews and original references), which was
exploited for gluelump Green's function in \cite{21}
\be
G_{\mu\nu} (x,y) = tr_a \int^\infty_0 ds  (Dz)_{xy} e^{-K}\lan
W^{(F)}_{\mu\nu} (C_{xy})\ran\label{16}\ee where $K= \frac14
\int^s_0 \left( \frac{dz_\mu}{d\tau}\right)^2 d\tau$ and
\be
W^F_{\mu\nu}(C_{xy}) =PP_F\left\{\exp (ig \int A_\lambda
dz_\lambda ) \exp (2 ig \int^s_0 d\tau \hat F_{\sigma \rho}
(z(\tau)))\right\}_{\mu\nu}\label{17}\ee
 and the closed contour $C_{xy}$ is formed by the straight line
 from $y$ to $x$ due to the heavy adjoint source Green's function
 and the path of the valence gluon $a_\mu$ from $x$ to $y$.
 Note that the nontrivial $\{\mu\nu\}$ dependence of the r.h.s. of
 (15) occurs only due to the $\hat F_{\nu\rho}$, expanding in
 powers of this term, one has $W_{\mu\nu}^F= W^{(0)}
 \delta_{\mu\nu} + W^{(1)}\delta_{\mu\nu} + W^{(1)}\hat F_{\mu\nu}
 +...$
 Neglecting in $G_{\mu\nu}$ gluon fields altogether we obtain the
 perturbative result, $G_{\mu\nu}  \to G^{(0)}_{\mu\nu}$
 \be
 G_{\mu\nu}^{(0)} (x,y) =\frac{N_c(N^2_c-1)
 \delta_{\mu\nu}}{4\pi^2(x-y)^2},~~ f^{(0)} = \frac{N_c(N_c^2-1)}{4\pi^2(x-y)^2}.\label{18}\ee
This is the leading term in the expansion of $G_{\mu\nu}$ at small
$|x-y|$, while the higher order  terms are given by the  OPE
formalism \cite{25}
\be
G_{\mu\nu} (x,y) =  G^{(0)}_{\mu\nu} (x,y) (1+C_1\alpha_s \ln
|x-y| +...)+ \delta_{\mu\nu} ( C_0 (x-y)^2 +C_4(x-y)^4+...)
\label{19}\ee where it is always assumed that $ C_0\sim D_1(0)
\sim \lan tr F^2_{\mu\nu} (0)\ran$ is finite.  The analysis in
\cite{26} using also nonperturbative nonlocal operators is
supporting the expansion (\ref{19}). To test the behaviour of
$G_{ik} (x,y)$ at small $(x-y)^2$, one should take into account
that small Wilson loops have a typical limiting form
\cite{1},\footnote{We do not discuss here the renormalization
properties of the r.h.s. of (18), since as it is seen in (22),
those reduce to the renormalization of gluonic condensate, widely
discussed in the framework of the OPE and ITEP sum rules [25].}
\be\lan W\ran = \exp \left( -\frac{g^2\lan tr F^2(0)\ran}{24
N_c}S^2\right)\label{18b}\ee where $S$ is the minimal area of the
small loop,
\be
S^2=\int_C z_\mu (x) dx_\nu \int_C z_\mu (x') dx'_\nu \approx \int
z^2_\mu (t) dt T\label{19b}\ee where $T\equiv |x-y|$. Introducing
(\ref{18b}) into (14) one obtains the path integral representation
\be
G_{ik} (x,y) =\delta_{ik} \int^\infty_0 ds (Dz)_{xy} \exp \left\{
-\frac14 \int^s_0 \dot{z}^2_\mu (\tau) d\tau -\omega^2\int^s_0
z^2_\mu d\tau\right\}\label{20b}\ee which can be estimated at
small  $|x-y|$, as it is explained in Appendix 1. The result is
\be
G_{ik} (x,y) |_{|x-y|\to 0} \approx \frac{N_c(N_c^2-1)}{4\pi^2
T^2} \left( 1- \frac{\omega^2 T^2}{4} +...\right)\label{21b}\ee
with $\omega^2=\frac{g^2}{12 N_c} \lan tr F^2\ran T^2$, which
yields for $D_1$ according to (\ref{13}),
\be
D_1(x) = \frac{4 C_2\alpha_s}{\pi} \left\{ \frac{1}{x^4} +
\frac{\pi^2G_2}{24 N_c} +...\right\}, \label{22b}\ee where $G_2$
is the standard gluonic condensate \cite{25}
\be
D(0)+D_1(0) =\frac{g^2}{12 N_c} tr F^2(0) =\frac{\pi^2}{18}
G_2.\label{23b}\ee One can check  consistency of the  resulting
$D_1(x)$.  First one  considers the singular term, $D_1^{sing} (x)
=\frac{4C_2\alpha_s}{\pi x^4}$ and inserts it in the static $Q\bar
Q$ potential. The static $Q\bar Q$ potential can be expressed
through $D$ and $D_1$, as was done in \cite{27}. \be V(r) =2r
\int^r_0 d\lambda \int^\infty_0 d\nu D(\lambda, \nu) + \int^r_0
\lambda d \lambda \int^\infty_0 d\nu [-2 D(\lambda, \nu) + D_1
(\lambda, \nu) ]\label{24b}\ee

Inserting in (\ref{24b}) the perturbative part of $D_1$ from
(\ref{22b}) one obtains the standard color Coulomb potential
$V(r)=- \frac{4\alpha_s}{3 r}$, thus checking the correct
normalization of $D_1(x)$.

Coming now to the constant term in (\ref{22b}) one can compare
$D_1(0)$ on the l.h.s. of (\ref{22b}) with the r.h.s., $ D_1(0) =
\frac{\alpha_s C_2}{\pi}\cdot \frac{\pi^2}{18} G_2 =
\frac{\alpha_s C_2}{\pi} (D(0)+D_1(0))$, where $D(0)+D_1(0)$ on
the r.h.s. of (\ref{22b}) are defined by the gluon condensate, Eq.
(\ref{23b}). Since $\frac{ \alpha_s C_2}{\pi} \la 1$, this
estimate of $D_1(0)$ is reasonable and suggests that for $\alpha_s
=0.4$ the magnitude of $D_1(0)$ is 0.2 $D(0)$.

This ratio is in agreement with the lattice calculations in
\cite{6}.

Another form of $G_{ik} (x,y)$ is available at all distances and
practically important at large $|x-y|$, namely
\be
G_{ik} (x,y) = N_c (N^2_c-1)\sum^\infty_{n=0} \Psi^{(i)}_n(0)
\Psi^{(k)^+}_n (0) e^{-M_n|x-y|}\label{20}\ee where
$\Psi_n^{(i)}(x), M_n$ are  eigenfunction and eigenvalue of the
gluelump Hamiltonian,  which is derived  using  the FFS path
integral  \cite{21}, details are given in Appendices 1,2,3,4 for
the convenience of the reader. \be H^{(glump)} =H_0^{(glump)}+
H_{spin} + H_C.\label{21} \ee Omitting the spin-splitting term
$H_{spin}$ and the pertubative gluon exchange term $H_C$, which
provide small corrections to the main term, one has for
$H_0^{(glump)}$ \cite{21}
\be
H_0^{(glump)} =\frac{\mu}{2} + \frac{p^2_r}{2\mu}
+\frac{L(L+1)/r^2}{2(\mu+\int^1_0 d\beta \beta^2
\nu(\beta))}+\frac{\sigma^2_{adj} r^2}{2} \int^1_0
\frac{d\beta}{\nu(\beta)}+\int^1_0\frac{\nu(\beta)}{2}
d\beta.\label{22} \ee Here $\mu,\nu$ are the so-called einbein
functions to be found exactly from the stationary point of the
Hamiltonian, $\frac{\delta H}{\delta \mu} =\frac{\delta
H}{\delta\nu} =0$, or approximately \cite{3,27} from the
stationary point of eigenvalues, e.g. $\frac{\delta
M_0}{\delta\mu}=\frac{\delta M_0}{\delta\nu} =0$. In what follows
we shall be interested in the case $L=0$, when $H_0$ reduces to
\be H_0^{(glump)} =\frac{\mu}{2}+\frac{p^2_r}{2\mu} +\sigma_{adj}
r\to \sqrt{\vep^2}+\sigma_{adj}r.\label{23}\ee For
$\Psi_n^{(\mu)}(0)$ one can use the known  equation \cite{28},
which is obtained from the eigenfunctions $\Psi_n$ of $H_0$
through the connection \cite{29} \be
 \Psi^{(\mu)}_n=
\frac{e_\mu}{\sqrt{2\mu}}\psi_n,~~ \left
(\Psi_n^{(\mu)}(0)\right)^2=\frac{\sigma_{adj}}{4\pi}.\label{24}
\ee

Inserting (\ref{24}) into (\ref{20}) one obtains
\be
G_{\mu\nu} (x,y)\approx  N_c(N^2_c-1) \sum^\infty_{n=0}
\delta_{\mu\nu}\frac{(\sigma_{adj})}{4\pi}e^{-M_n|x-y|}.\label{25}\ee
It is clear that for $x\to y$ the sum in (\ref{25}) diverges and
one should use instead of (\ref{25}) the perturbative answer
(\ref{18}). For large $|x-y|$ one can keep in (\ref{25}) only the
terms with the lowest mass, i.e. for the color electric gluelump
state $1^{--}$, which obtains for spacial $\mu,\nu=i,k$

Thus one gets \be G_{ik}\left|_{|x-y|\to \infty} \right. \approx
(N^2_c-1)\frac{N_c\sigma_{adj}}{4\pi} \delta_{ik}
e^{-M_0|x-y|}.\label{26}\ee The eigenvalue $M_0$ was found in
\cite{21} to be $M_0\cong1.5$ GeV for $\sigma_f=0.18 $ GeV$^2$.

Using (\ref{13}) one can define from (\ref{26}) the
nonperturbative part  of $D_1$, which is valid at large $|x|$, \be
D_1^{(nonp)} (x) = \frac{C_2(f)\alpha_s 2M_0
\sigma_{adj}}{\sqrt{x^2}} e^{-M_0|x|},~~ C_2(f)
=\frac{N^2_c-1}{2N_c}\label{32}\ee and the total $D_1$ due to
(\ref{13}) and (\ref{18}) can be represented as \be D_1(x)
=\frac{4C_2(f)\alpha_s}{\pi x^4}+ O(\alpha^2_s) + D_1^{(nonp)}
(x)\label{33}\ee

\section{The correlator $D(x)$}

We now turn to the function $D(x)$ and to this end we specify the
indices fo $D_{\mu\nu,\lambda\sigma}$ in (\ref{6}) as $D_{ik, lm}
(x,y)$ with $i,k,l,m=1,2,3$ and take the interval
$h_\mu=x_\mu-y_\mu$ to lie on the temporal axis,  $h_4\neq 0,
h_i\equiv 0, i=1,2,3$.  One can again represent $D_{ik, lm}$ as in
(\ref{6}), \be D_{ik lm}(x,y) = D^{(0)}_{ik,lm} + D^{(1)}_{ik,lm}+
D^{(2)}_{ik,lm}\label{34}\ee

 As in the previous section, $D^{(0)}$ contributes to the function
 $D_1$, while the contour differention operation
 $\overleftarrow{\delta}_{\mu}$  introduces the new field
 $F_{\mu4}$, so that one has to do with the triple correlator
 $tr\lan F\Phi F\Phi F\ran$. In this paper we are considering the
 Gausian approximation for simplicity and neglect all correlators
 except for the quadratic ones, Eq. (\ref{1}). Therefore the term
 $D^{(0)} $ and $D^{(1)}$ in
 (\ref{34}) do not contribute to $D(x)$, and we concentrate on the
 last term in (\ref{34}), $D^{(2)}$, which can be written as
 \be D^{(2)}_{ik, lm} (x,y) =- \frac{g^4}{2N_c^2}\lan tr_a ([a_i,
 a_k]\hat \Phi (x,y) [a_l,a_m])\ran.\label{35}\ee

 This function can be connected to the two-gluon gluelump Green's
 function.
The two gluon gluelumps considered in \cite{21}  belong to the
symmetric in color and spin components, while here one can rewrite
(35) as \be [a_{i}, a_k] = i  a^a_i a^b_k f^{abc} T^c\label{36}\ee
and the resulting gluelump function is \be G_{ik,lm} =tr_a\lan
f^{abc} f^{def} a_i^a(x) a_k^b(x) T^c\hat \Phi(x,y) T^fa^d_l
a^e_m\ran.\label{37}\ee  One can immediately see that the gluelump
in (\ref{37})  is antisymmetric both in color and spin indices,
but the total  wave function is symmetric and the relative angular
momentum $L$ of the lowest state can be taken as $L=0$. One can
fix in (\ref{37}) color indices $a,b;d, e$ and average the Green's
function over all fields $A_\mu^h$ with $h\neq a,b;d, e$. Using
the same argument as it was done for the one-gluon gluelump
function, one can argue that this averaging will produce the white
string (of triangle shape at any given moment), and hence it will
ensure terms $(\delta_{ad} \delta_{be} +$ permutations). As a
result one can represent $G_{ik, lm}$ in the form \be G_{ik,
lm}(x,y) = N^2_c (N^2_c-1) (\delta_{il} \delta_{km}-\delta_{im}
\delta_{kl}) G^{(2gl)}(x,y)\label{38}\ee where $G^{(2gl)}(x,y)$ is
the Green's function of the two-gluon gluelump, which was studied
in \cite{21}, and the Hamiltonian and lowest eigenvalue was found
there explicitly (see also Appendix A of \cite{21}).

Comparison of  Eqs. (\ref{1}), (\ref{35}) and (\ref{38})
immediately yields the following expression for $D(x)$.
\be
D(x-y) =\frac{g^4 (N_c^2-1)}{2} G^{(2gl)} (x,y).\label{39}\ee If
one exploits for $G^{(2gl)} $ the perturbative expression
\be
G^{(2gl)(0)} (x,y) = \frac{1}{(4\pi^2(x-y)^2)^2} +
O\frac{(\alpha_s\ln (x-y))}{(x-y)^4}\label{40}\ee then one
recovers the corresponding perturbative expansion for $D(x)$,
which was studied in \cite{29}. However, as it was shown in
\cite{30}, all perturbative terms of $D(x)$ are cancelled by those
of higher correlators, so that they do not contribute the
(divergent) terms in the expression for the string tension
\be\sigma=\frac12 \int d^2 x
(D(x)+higher~correlators).\label{41}\ee

Nonperturbative contribution to $D(x)$ can be written down using
the spectral decomposition for $G^{(2gl)} (T), T=x_4-y_4,$
\be
G^{(2gl)} (T) =\sum |\Psi_n^{(2gl)} (0)|^2 e^{-M_n^{(2gl)}
T}.\label{42}\ee

Here $\Psi_n^{(2gl)} (\vexi, \veta)$ is the two-gluon gluelump
wave function calculated with the string Hamiltonian,  neglecting
spin-spin interaction in the first approximation, considered in
\cite{21}, $\vexi, \veta$ are Jacobi coordinates in the system of
two gluons and the adjoint fixed center.

The calculation of $M_n^{(2gl)}$ and $\Psi^{(2gl)}_n (0)$ is
discussed in the Appendix 2, and here we quote the final result
for the lowest gluelump state:
\be
M_0^{(2gl)} =2.56 {\rm GeV},~~ | \Psi_n^{(2gl)}(0)|^2=0.108
\sigma^2_f.\label{43}\ee Hence the leading at large $T$
asymptotics of $D(x)$ is \be D^{(2gl)}(x) \cong
\frac{g^4(N^2_c-1)}{2} 0.108 \sigma^2_f e^{-M_0^{(2gl)}|x|},~~
M_0^{(2gl)}|x|\gg 1, \label{44}\ee

The corresponding value of the gluon correlation length is very
small $T_g^{(2g)} = \frac{1}{M_0^{(2gl)} } \cong 0.08$ fm, and
different from the correlation length of $D_1(x)$, which is
$T_g^{(1g)} \equiv \frac{1}{M_0^{(1gl)}}=0.13$ fm. This fact is in
contradiction with the lattice calculations of the Pisa group  in
\cite{6}, where both correlation lengths coincide. To understand
the reason for this discrepancy one must consider the higher in
$\alpha_s$ terms contributing to $D(x)$. Indeed, using the term
$L_3$ in the Lagrangian, which transforms two-gluon state into
one-gluon and three-gluon states, one obtains in $D(x)$ the
one-gluon state with the same asymptotics as in $D_1(x)$, namely
\be
D(x)= D^{(2gl)}(x) + C_1\alpha_s^3 D^{(1gl)} (x) + C_2\alpha_s^3
D^{(3gl)}(x)+...\label{45}\ee Here $D^{(1gl)}(x)\sim \exp
(-M_0^{(1gl)} |x|),~~ D^{(3gl)}(x) \sim \exp (-M_0^{(3gl)}|x|)$.
In a similar way $D_1(x)$ acquires terms with two-gluon and
three-gluon gluelump asymptotics. Still effectively one expects
that correlation lengths satisfy $T_g(D_1)> T_g (D)$, and this in
agreement with lattice calculations of Bali et al (last ref. in
{\cite{6}).

\section{Discussion of results}

In (32), (33) and (\ref{44}) we have obtained the perturbative and
nonperturbative parts of $D_1(x)$ and $D(x)$ and this is the main
result of the paper.

Coming back to the  three basic points outlined in the
Introduction, the first point refers to the distinction between
perturbative and nonperturbative and the essence of the
nonperturbative mechanism of confinement. In this regard one
should consider the mechanism which creates massive gluelump,
namely  via the path-integral representation (\ref{16}) and using
Wilson loop area law  one obtains the Hamiltonian  (27), where
nonperturbative dynamics is connected to the string tension
$\sigma$. The latter in its turn is expressed through $D(x)$ as in
(\ref{41}). Thus the signature for the nonperturbative is
$\sigma\neq 0$, and one can separate from $D(x), D_1(x)$
perturbative parts as the limit of $D(x)$, $D_1(x)$ when $\sigma$
tends to zero.

One should stress here that it would be in general incorrect to
represent $D(x)$, $D_1(x)$ as a sum of perturbative and
nonperturbative components. The expansion (45) gives instead
another type of representation where the perturbative and
nonperturbative are strongly mixed up. One can only state, that at
small distances correlators behave perturbatively, while at large
distances decay nonperturbatively.

 At this point it is necessary
to ask, what is the nonperturbative mechanism which creates $D(x)$
and according to (\ref{41}) also creates confinement itself.

The answer lies in our separation of (\ref{3}), where in contrast
to the usual background formalism \cite{22}, both fields $a_\mu^a$
and $B_\mu^b$ are intrinsically the same, but the fields $B^b_\mu$
with color index $b\neq a$ are assumed to form the collective
field which after averaging acts as a "white force" on the gluon
$a^a_\mu$ with the correlators $D(x), D_1(x)$ as a measure of this
force. As a result the gluon $a^a_\mu$ forms the bound state,
which can be used to calculate $D(x), D_1(x)$. In this way the
problem becomes self-consistent, when one can ensure that the same
$D(x), D_1(x)$ act as input in the "white force"and result in the
outcome of calculation of correlators. In the  present paper this
selfconsistency was checked only partially on the level of the
string tension $\sigma$.

At the same time the selfconsistency allows one to connect
$\Lambda_{QCD}$ and $\sigma$, thus reducing the number of
parameters in theory to one, as it should be in QCD.

Indeed, calculating $\sigma_f$ via (\ref{41}) and using (\ref{44})
for $D(x)$ one obtains
\be
\sigma_f =\left( \frac{\alpha_s}{0.3}\right)^2 0.53
\sigma_f\label{46}\ee where $\sigma_f$ on the r.h.s. is coming
from the input $D(x)$, while that on the l.h.s. is from the
resulting $D(x)$. For reasonable values of $\alpha_s\sim 0.41$ one
obtains that $D^{(2gl)}(x)$ satisfies the selfconsistency
criterium (\ref{46}).  Neglecting all other contributions to
$D(x)$ in (\ref{45}), one can consider (\ref{46}) as an equation
defining the connection  between $\sigma_f$ and  $\Lambda_{QCD}$.
Namely writing $\alpha_s =\alpha_s (M_0^{(2gl)}) =0.41=
\frac{4\pi}{\beta_{0}ln \left(
\frac{M_0^{(2gl)}}{\Lambda_{QCD}}\right)^2}$, one finds
 \be
\Lambda_{QCD} (n_f=0) = 0.25 M_0^{(2gl)} = 0.88
\sqrt{\sigma_f},\label{47a}\ee and for $\sigma_f =0.18 $ GeV$^2$
one has $\Lambda_{QCD} =0.375$ GeV. One should compare this result
with the value of $\Lambda_{QCD}^{( \overline{MS})} $, obtained on
the lattice (see e.g. [32] and refs. there),  $\Lambda_{QCD}^{(
\overline{MS})}  (n_f=0) =0.242$ GeV, which is easily recalculated
for $\Lambda^{(V)}_{QCD} = (385 \pm 30)$ MeV, in good agreement
with our theoretical value.

Keeping in $D(x)$ also higher order terms, as in (45) and
inserting in  (41) one obtains instead of (46) an expansion \be
\sigma _f = \left( \frac{\alpha_s}{0.41}\right)^2 \sigma_f + C_1
\alpha_s^3 \sigma_f +..\label{48aa}\ee In this way one finds a
general type of connection between $\sigma_f$ and $\Lambda_{QCD}$,
as was stated in the point 3 of Introduction. Coming now to the
small $x$ behaviour in (\ref{44}) it should be replaced by an
analytic one, $D(x) \sim c_0 + c_1 x^2+...$, as it follows from
\cite{31}, and the gluonic condensate calculated from
$D^{(2gl}(0)$ in (\ref{44}), which is much larger than the
standard value \cite{25}, will be  replaced by a smaller value,
and in addition also $D^{(1gl)} (x)$ in (\ref{45}) contributes to
$\sigma_f$. Therefore a more detailed analysis, including the
behaviour of $D(x)$ at small $x$ is required, which is relegated
to future publications.

Summarizing the results, we have computed the first terms of
perturbative and nonperturbative field correlators $D(x)$ and
$D_1(x)$ in the expansion in powers of $\alpha_s$ with coefficient
functions proportional to the gluelump Green's function, and have
made the first check of selfconsistency of the resulting string
tension. In this way the preliminary analysis in the paper
supports the confinement mechanism as the formation of the
selfconsistent background field acting as a white string on
propagating gluons.

The author is grateful to A.M.Badalian and Yu.S.Kalashnikova  for
fruitful discussions, and V.I.Shevchenko for collaboration at the
first stages of the work.

 The work  is supported
  by the Federal Program of the Russian Ministry of industry, Science and Technology
  No.40.052.1.1.1112, and by the
grant for scientific schools NS-1774. 2003. 2.\\

\vspace{2cm}

{\bf Appendix 1}\\

{\bf  Derivation  of Eq. (7) }\\

 \setcounter{equation}{0} \def\theequation{A1.\arabic{equation}}

Consider the general bilocal correlator, \be \Psi(x, y) \equiv
\lan tr K_1 (x) \hat \Phi (x,y) K_2(y)\ran \label{a1.1}\ee where
$\hat \Phi(x,y)$ is the adjoint parallel transporter,
\be
\hat \Phi(x,y) = P\exp ig \int^x_y \hat A_\mu d
z_\mu\label{a1.2}\ee

Taking the difference
\be
\Delta_\mu \Psi (x,y) \equiv \Psi (x+\delta x_\mu, y) = \lan tr
K_1 (x+\delta x_\mu) \hat \phi(x+\delta x_\mu,y)
K_2(y)\ran-\Psi(x,y),\label{a1.3}\ee one can deform identically
the shifted contour by insertion of double lines of the $L$ form
$N$ times at distance $a$ from each other,  $Na = |x-y|$. The
explicit  form of deformation is shown in Fig.2 of  the last ref.
[23]. In the limit when $\delta x_\mu\to 0,$ and $a\to 0$, $N\to
\infty$, any open ($a\times a)$ plaquette from two neighboring $L$
insertions tends to $(F_{\mu\lambda} a^2)$ and  one obtains
\be
\frac{\partial_\mu \Psi(x,y)}{\partial x_\mu} = \lan tr \left (
\frac{\partial K_1(x)}{\partial x_\mu} - ig  K_1(x) \hat A_\mu (x)
\right) \hat \Phi(x,y) K_2(y)\ran + \lan tr K_1 (x) \delta_\mu (x)
\hat \Phi(x,y) K_2(y)\ran\label{a1.4}\ee where $\delta_\mu(x)$ is
defined in Eq.(8).

\vspace{2cm}

{\bf Appendix 2}\\

{\bf Calculation of $D_1(x)$ at small $x$ }\\

 \setcounter{equation}{0} \def\theequation{A2.\arabic{equation}}

One must calculate the path integral for the gluon Green's
function
\be
G_{ik}{(x,y)} = \delta_{ik} \int ds (Dz)_{xy} e^{-K}\lan
W\ran\label{1.1} \ee where $\lan W\ran$ is given in (\ref{18b})
and $K=\frac14 \int^s_0 \dot{z}^2_\mu(\tau) d\tau$.

Making an  approximation (\ref{19b}) and introducing the dynamical
mass variable $\mu(t) = \frac{dt}{2d\tau}$, where $\tau\leq s$ in
the proper time \cite{2,3}, one can replace $ds(Dz_4)_{x_4, y_4}$
by the path integral $D\mu$ see \cite{35} for more details) as
follows
\be
G_{ik}(0T)=\delta_{ik} \int \frac{1}{2\bar \mu}(D^3 z)_{0,0} D\mu
e^{-\int^T_0 dt \left( \frac{\mu}{2}+ \frac{\mu\dot{\vez}^2}{2}
+\frac{\mu\omega^2\vez^2}{2}\right)}\label{1.2} \ee where we have
defined $$ \frac{\mu\omega^2}{2}=\frac{g^2}{24 N_c} \lan tr
F^2_{\mu\nu}\ran T, (D^3 z) D\mu = \prod^N_{n=1} \frac{d^3\Delta
z(n)}{l^3(n)} \frac{d\mu(n)}{l_\mu(n)},$$ and $$ l_\mu(n) = \left(
\frac{2\pi\mu(n)}{\Delta t} \right)^{1/2},~~ l(n) =
\left(\frac{2\pi\Delta t}{\mu(n)}\right)^{1/2}, ~~ \Delta t N =T,
~~ N\to \infty.$$ $$ \bar \mu = \frac{1}{s} \int^s_0 \mu (\tau)
d\tau \approx \frac{1}{T} \int^T_0\mu(t) dt.$$

The integration over $(D^3z)_{00}$ in (\ref{1.2}) is known from
the textbook solution for the Green's function of oscillator,
hence one has \be G_{ik} (0T)= \delta_{ik} \int \frac{D\mu}{2\bar
\mu} e^{-\int^T_0 dt \frac{\mu}{2}} \left( \frac{\bar \mu
\omega}{2\pi sh \omega T}\right)^{3/2} = \frac{1}{2\bar \mu}
\left( \frac{\bar \mu \omega}{2\pi sh \omega
T}\right)^{3/2}\label{1.3} \ee

Now one can obtain $\bar \mu$ from the solution of the free
Green's function at small $T$, which is realized when $\omega \to
0$, \be \frac{1}{2\bar \mu} \left( \frac{\bar \mu}{2\pi t}
\right)^{3/2} = \frac{1}{4\pi^2 T^2}, ~~ \bar \mu = \frac{2}{\pi
T}\label{1.4}\ee and expanding further (\ref{1.3}), one finally
obtains
\be
G_{ik} = \frac{1}{4\pi^2 T^2}\left( 1-\frac{\omega^2}{4}
T^2\right)\label{1.5}\ee and using relation (\ref{13}) one finds
resulting equation (\ref{22b}) for $D_1(x)$, where $$G_2\equiv
\frac{2\alpha_s}{\pi} \lan tr F^2_{\mu\nu}\ran.$$

\vspace{2cm}

{\bf Appendix 3}\\

{\bf  Derivation  of the gluelump Hamiltonian }\\

 \setcounter{equation}{0} \def\theequation{A3.\arabic{equation}}

One starts with the expressions (\ref{16}), (\ref{17}) for  the
gluelump Green's function, obtained in the framework of the FFS
formalism \cite{24}. The first step is the vacuum averaging of
$\lan W^F_{\mu\nu}\ran$, which is done using the nonabelian Stokes
theorem and the cluster expansion where we systematically keep the
lowest order (quadratic) terms in powers of field correlators, as
is explained in the Introduction. The result is
\be
\lan W_{\mu\nu}^F\ran= \exp \left\{ - \frac{\xi}{2} \int
d\rho_{\rho\lambda} (u,\tau)  \int d \rho_{\rho'\lambda'}
(u',\tau') D_{\rho\lambda, \rho'\lambda'}
(u,u')\right\}\label{A1.1}\ee where  we have defined
\be
\xi\equiv \frac{C_2(adj)}{C_2(f)}, ~~ d\rho_{\rho\lambda} (u,\tau)
= d\sigma_{\rho\lambda} (u) - 2 i  \hat S_{\rho\lambda}
d\tau,\label{A1.2}\ee and $d\sigma_{\rho\lambda} (u)$ is the
surface element, while $\hat S_{\rho\lambda}$ contains gluon spin
operators and prescription  of replacing the  coordinate $u$ of
the surface element by the coordinate of the boundary -- gluon
trajectory $z_{\mu} (\tau)$, \be \hat S_{\rho\lambda}
F_{\rho\lambda} (u) = (S_i H_i + \tilde S_i E_i)
\delta_{u,z(\tau)}\label{A1.3}\ee

In what follows we shall not use the explicit form of $S_i, \tilde
S_i$.

In (\ref{A1.1}) three types of terms are present:
\begin{description}
  \item[a] ~~~terms quadratic in $d\sigma$, which yield the area law for large area, much larger than
  $T^2_g$;
  \item[b] ~~~mixed term, proportional to $d\sigma_{\rho\lambda}
  \tilde S_{\rho\lambda}$. These  produce spin-dependent terms in
  the Hamiltonian, as it is explained in \cite{21}, [27].
  \item[c] ~~~terms quadratic in $\hat S$; they give rise to the
  self-energy terms of the propagating  gluon and are treated in
  Appendix 3. It is shown there, that  the self-energy terms
  occurring due to both quartic terms $a^4$ in the QCD Lagrangian
  and terms quadratic in $\hat S$ should cancel in the framework
  of BFF.

\end{description}
 As a result, if one neglects spin-dependent terms in the first
 approximation, one can keep in (\ref{A1.1}), (\ref{A1.2}) only the
 surface element, $d\rho\to d
\sigma$, and the r.h..s. of (\ref{A1.1}) simplifies, namely \be
\lan W^F_{\mu\nu}\ran = \delta_{\mu\nu} \exp (-\sigma_{adj}
A)\label{A1.4}\ee  where $A$ is the area between the gluon path
$z(\tau), \tau \epsilon [0,s],$ and the piece of the straight line
between $x$ and $y$. It is usually assumed on physical grounds
(and supported  by lattice data) that $A$ is the minimal area for
the given boundary.

As a next step one should write the resulting expression for the
gluelump Green's function, with gluon spin effects neglected, \be
G_{\mu\nu}^{(0)} (x,y) =\delta_{\mu\nu} \int^\infty_0 ds (Dz)_{xy}
e^{-K-\sigma_{adj} A},~~ K=\frac14 \int^s_0 d\tau (\dot z_\mu)^2
\label{A1.5}\ee in the form, where real time $t$ instead of the
proper time $\tau, s$ enters.

This is done in Appendix2, Eq. (A2.2) where the relation is used
$ds(D^4z)_{xy} = (D^3z) \frac{D\mu}{2\bar \mu}$, and explicit
expressions for $\bar \mu, D\mu$ are given. Here $\mu(t)$ connects
real (Euclidean) time $t$ and  proper  time $\tau,
2\mu(t)=\frac{dt}{d\tau}$.The transformations discussed below have
been introduced in \cite{Dub}.

One can write the  area $A$ in the Nambu-Goto form,
\be
A=\int^1_0 d\beta \int^T_0 dt \sqrt{\dot w^2_i w_k^{\prime 2}-
(\dot w_i w'_i)^2},~~ T=|x_4-y_4|, \label{A2.6}\ee with
$w(\beta,t)$ being the coordinate on the string world sheet, $\dot
w_i \equiv \frac{\partial w_i}{\partial t},~~ w'_i\equiv
\frac{\partial w_i}{\partial\beta}$.

One can  approximate the minimal area choosing $w_i(\beta)$ in the
form of the piece of the straight line, $w_i
(\beta,t)=r_i(t)\beta$.

One may wonder what is the accuracy  of the straight-line
approximation, i.e. of the neglecting the string excitation, which
in the background field formalism amounts to the hybrid
excitations, with the energy gap of around 1 GeV. Therefore all
nonadiabatic effects of this type are typically of the order of
the hybrid admixture to the given gluelump state. For the ground
state this admixture  was calculated and discussed in \cite{Sim},
and appeared to be of the order of few percent.

As a next step one introduces einbein variables to get rid of the
square root in (A2.6), and as a result the total Euclidean action
$ \mathcal{E}$ as a function of einbein parameters $\mu(t),
\eta(t), \nu(t)$ has  the form
\be
\mathcal{E}(\mu,\eta, \nu) =\int^T_0 dt \int^1_0 d\beta \left[
\frac{\mu\dot{r}^2_i}{2} + \frac{1}{2\nu} (\dot{w}^2_i
+(\sigma_{adj}\nu)^2 r^2 -2\eta
(\dot{w}_ir_i)+\eta^2r^2)\right]\label{A2.7}\ee Finally the
gluelump Green's function can be written as

\be
G_{\mu\nu}^{(0)} (x,y) =\delta_{\mu\nu} \int D^3 r_i D\nu D\eta
\frac{D\mu}{2\bar\mu} e^{ - \mathcal{E}}\label{A2.8}\ee

From (A2.8) integrating over $D\eta$ one can go over to the
Minkowskian Hamiltonian in the standard way using the Feynman
prescription:
\be
\int (Dx)_{if} e^{i action} = \lan i|e^{-iHT}|f\ran
\label{A2.9}\ee

The Hamiltonian $H=H(\nu,\mu)$ depends on the auxiliary einbein
functions $\nu,\mu$ and the remaining integral over $D\nu D\mu$
can be taken using  the stationary point method. The gluelump
Hamiltonian $H(\nu,\mu)$ has the form \cite{21}, given in Eq.
({27}) of the main text
\be
H^{(glump)} =\frac{\mu}{2} + \frac{p^2_r}{2\mu} +
\frac{L(L+1)}{2Ir^2}+ \frac{(\sigma_{adj}r)^2}{2} \int^1_0
\frac{d\beta}{\nu(\beta)} + \int^1_0 \frac{\nu(\beta)}{2} d\beta
\label{A2.10}\ee where the  factor of inertia moment $I$ is
\be
I=\mu+ \int^1_0 d\beta \beta^2 \nu (\beta)\label{A2.11}\ee

For $L=0$ the Hamiltonian is easily computed taking stationary
point values of $\nu=\nu_0 = \sigma_{adj} r$ and $\mu=\mu_0 =
\sqrt{p^2_r}$, \be H(L=0)=\sqrt{p^2_r} +
\sigma_{adj}r\label{A2.12}\ee

The lowest eigenvalue of (A3.12) is easily calculated to be
$M_0\approx 1.5$ GeV. Note  that perturbative  interaction is not
included in (A3.12), taking that into account the eigenvalue drops
to $M_0 \approx 1$ GeV for $\alpha_s(r)\approx \bar \alpha_s
\approx 0.2$. Since this interaction is subject to strong
radiative corrections of destructive character,  we do not take it
into account, see \cite{21} for more discussion.

\vspace{2cm}

{\bf Appendix 4}\\

{\bf  The gluon self-energy and auxiliary function method for the
effective Lagrangian of QCD }\\

 \setcounter{equation}{0} \def\theequation{A4.\arabic{equation}}

One starts with the decomposition of the gauge field $A_\mu$ into
background component $B_\mu$ and valence gluon component $a_\mu$,
\be  A_\mu =B_\mu +a_\mu.\label{au1}\ee

We shall not specify below the principle of decomposition, one can
use e.g. the idea of separating one or a group of few colors for
$a_\mu$,  while the rest of colors is in $B_\mu$, as it was done
in \cite{17,36}, which assumes a fixed gauge for $a_\mu$. The
following derivations are of general character and do not depend
on the assumed  separation. Writing  the QCD Lagrangian as
\be
L(A) =L(B+a)= \frac12 tr F^2_{\mu\nu} (B+a),\label{au2}\ee where
$F_{\mu\nu}$ is $$ F_{\mu\nu} (x) =\partial_\mu A_\nu-\partial_\nu
A_\mu-ig [A_\mu,A_\nu]=$$ \be=
\partial_\mu(a_\nu+B_\nu)-\partial_\nu(a_\mu+B_\mu)-ig[a_\mu+B_\mu,a_\nu+B_\nu]=\label{au3}
\ee $$=\hat D_\mu a_\nu-\hat D_\nu a_\mu-ig[a_\mu,a_\nu]+
F^{(B)}_{\mu\nu}$$ one obtains the effective action for the field
$a_\mu$, $S_{eff}(a)$,  after integrating out  the  fields
$B_\mu$, namely
 \be
 \lan e^{-\int L(B+a) d^4 x}\ran_B\equiv e^{-S_{eff} (a)}.
 \label{au4}
 \ee
 where
 \be
S_{eff} (a) =\int \lan L (B+a)\ran_B d^4 x -\frac12 \int d^4 x
\int d^4 y \llan L(x)L(y)\rran +... \label{au5} \ee

In what follows we shall be interested in  the  terms of the
fourth order in $a_\mu$, which appear in the path-integral
representation of the valence gluon Green's function, namely
\cite{24} \be G_{\mu\nu} \equiv \lan a_\mu(x) a_\nu
(y)\ran_{a,B}=\lan (\hat D^2_\lambda\delta_{\mu\nu}- 2ig \hat
F_{\mu\nu})^{-1}_{x,y} \ran_B\label{au6}\ee where the background
Feynman gauge is assumed for $a_\mu, D_\mu(B) a_\mu=0$.

Writing (\ref{au6}) as a path integral, one has
  \be
      G_{\mu\nu} (x,y) =
      \int^{\infty}_0 ds (Dz)_{xy} e^{-K_0}\Phi_{\mu\nu}(x,y)
      \label{au7}
      \ee
      where we have defined
      \be
      K_0=\frac14 \int^\infty_0 \left (\frac{dz_\mu}{d\tau}\right)^2
      d\tau;~~ \Phi_{\mu\nu} (x,y) =[P_FP_A\exp (ig \int^x_y A_\lambda
      dz_\lambda)\times\label{au8}
      \ee
      \be
      \times \exp (2g\int^s_0 d\tau F_{\sigma\rho}
      (z(\tau)))]_{\mu\nu}
      \label{au9}
      \ee
The last exponential factor in(\ref{au9}) represents the
interaction of the valence gluon color magnetic moment with the
background field, and the averaging of this factor over $B_\mu$
produces the effective self-energy term which is negative and
which makes the gluon unstable (mass eigenvalues are not bounded
from below). This is the paramagnetic instability of the
relativistic charge moving in the Gaussian stochastic environment.
A similar situation occurs for the quarks where the corresponding
term has the form $exp(g\sigma_{\mu\nu} \int^s_0 d\tau F_{\mu\nu}
(z(\tau)))$ and produces also the negative self-energy
contribution, not bounded from below when  considered outside of
perturbation theory. In the quark case however, the term is a
$4\times 4$ Dirac matrix and one recovers the stability when
considering also negative energy states (lower Dirac components)
-- see \cite{37} for a discussion.

In the gluon case this  mechanism is absent, and we now show that
there is another stabilyzing phenomenon, associated with the gluon
term $\sim a^4_\mu$ in the QCD Lagrangian.

To this end we consider two characteristic terms in the
Lagrangian: the original quartic and another effective quartic
generated by the paramagnetic term in $L(B+a)$: \be L^{(F)}
=gf^{ika} a_\mu^i a_\nu^k F^a_{\mu\nu} (B)\label{au10}\ee

After averaging the square of $L^{(F)}$ as in the second term in
(\ref{au5}), one obtains the effective quartic term in the
Lagrangian \be \lan e^{-L}\ran_{quartic} =  e^{-S_{eff}^{(4)}};~~
S_{eff}^{(4)} = \int d^4x d^4y a^a_\mu (x) a_\nu^b (x) a^{a'}_\mu
(y) a^{b'}_\nu (y) f^{abc} f^{a'b'c} K(x,y)\label{au11}\ee where
we have defined \be K(x,y) = \frac{g^2}{4} \delta^{(4)} (x-y)
-\frac{2N_c}{N_c^2-1}  D(x-y)\label{au12}\ee and $D(x)$ is the
standard field correlator \cite{1,2} \be
  g^2\llan F_{\lambda\beta}(x) F_{\gamma\delta}(y)\rran
  =(\delta_{\lambda\gamma}
  \delta_{\beta\delta}-\delta_{\lambda\delta}
  \delta_{\beta\gamma})D(x-y)+O(D_1).
  \label{au13}
  \ee

One can rewrite $S^{(4)}_{eff}$ more conveniently as  \be
S^{(4)}_{eff} =\int \Psi_{aa'}(x,y) \tilde K_{aa',bb'} (x,y)
\Psi_{bb'}(x,y) d^4 x d^4y \equiv \Psi \tilde K\Psi
\label{au14}\ee and use the Hubbard-Stratonovich identity
\be
e^{-\Psi\tilde K \Psi} = \int \sqrt{ det\tilde K} D\chi e^{-\chi
\tilde K\chi+i\Psi \tilde K \chi+i\chi\tilde
K\Psi}.\label{au15}\ee In (\ref{au14}), (\ref{au15}) we have used
notations:
\be
\tilde K_{aa',bb'} (x,y) =f^{abc} f^{a'b'c}K(x,y)\label{au16}\ee
\be \Psi_{aa'} (x,y) =a_{\mu}^a (x) a_\mu^{a'} (y)\label{au17}\ee
\be
\chi\equiv \chi_{ aa'} (x,y)\label{au18}\ee

On the r.h.s. of (\ref{au15}) the gluons $a_\mu$ enter only
quadratically; combining quadratic terms  from $L$ with the
latter, one obtains the term in the partition function, \be
Z=\int\exp (-S^{(2)}_{eff} - \chi \tilde K \chi ) D\chi
\label{au19}\ee with \be S^{(2)}_{eff} =\frac12 \int a_\mu^a (x)
[-\hat D^2_{ab} \delta_{\mu\nu} \delta(x-y) -2i I^{ab} (x,y) ]
a^b_\nu(y) dxdy\label{au20}\ee where we have defined \be
I^{aa'}(x,y) = \tilde K_{aa',bb'} (x,y) \chi_{bb'}
(x,y)\label{au21}\ee

After integrating out $Da_\mu$ one obtains an effective Lagrangian
for the auxiliary fields $\chi$, \be Z\sim \int D\chi e^{-L_{eff}
(\chi)},~~ L_{eff} (\chi) =\frac12 tr\ln (-D^2-2iI)+ \chi\tilde K
\chi.\label{au22}\ee

Using the stationary point method for the integral over $D\chi$,
one has equations
\be
\left.\frac{\delta L_{eff} (\chi)}{\delta \chi}
\right|_{\chi=\chi^{(0)}}=0, ~~ \chi^{(0)}=\frac{i}{2}
\frac{1}{(-D^2-2iI)}\label{au23}\ee and defining the effective
mass operator, \be \mathcal{M}^2_0\equiv -2iI =- 2i \tilde K
\chi^{(0)}\label{au24}\ee one has equation for $ \mathcal{M}^2_0$,
\be  \mathcal{M}^2_0 = \tilde K \frac{1}{-D^2+
\mathcal{M}^2_0}.\label{au25}\ee

A nonzero solution for $M^2_0$ in (25) would imply the existence
of the effective mass term of the valence gluon. At the same time
the ghost  Green's function does not have  both quartic and
magnetic moment contribution, so that  for the ghost the effective
mass is zero. To ensure the exact cancellation of the ghost and
unphysical gluon degrees of freedom one must  require that $
\mathcal{M}^2_0$  vanish.

The r.h.s.  of (A4.25) is ultraviolet divergent and needs
renormalization, which can be accomplished to satisfy this
requirement. Therefore we shall assume everywhere that the
effective gluon mass is zero and the gluon magnetic moment term
with  $\hat F_{\sigma\rho}(z(\tau))$ in Eq.(15) is absent, as it
was assumed in Appendix 3.

\vspace{2cm}

{\bf Appendix 5 }\\

{\bf Spectral representation of the three-body Green's function
}\\

 \setcounter{equation}{0} \def\theequation{A5.\arabic{equation}}

 In section 3 we have to evaluate the spectral representation of
 $G^{(2gl)} (T)$ in (\ref{42}) and the lowest eigenvalue
 $M_0^{(2gl)}$.  In this Appendix we  give some details of
 derivation, using Appendix A of \cite{21}.
  The total eigenfunction of two-gluon gluelump $\Psi(\vexi,
  \veta)$ where $\vexi, \veta$ are Jacobi coordinates,
\be
  \veta = \ver_{12}/\sqrt{2}, \vexi = (\ver_1 +
  \ver_2)/{\sqrt{2}}, ~~ \rho^2= \vexi^2+\veta^2\label{A.1}\ee
  can be expanded in the hyperspherical basis \cite{33}.
  \be \Psi=\sum_{K,\nu} u^\nu_K (\Omega) \frac{y^\nu_K
  (\rho)}{\rho}\label{A.2}\ee
  where $y^\nu_K(\rho)$ is the hyperradial wave function and the
  series (\ref{A.2}) is fast converging for linear confinement (for a
  recent review and refs. see \cite{34}), so that one can retain
  the lowest term in (\ref{A.1}) for the two-gluon gluelump
  corresponding to $K=0$.
  With the definition $u^0_0(\Omega) = \frac{1}{\sqrt{\pi^3}}, ~~
  y^0_0(\rho) \equiv y(\rho)$, the equation for $y(\rho)$ has the
  form
  \be y^{\prime\prime} + 3\frac{d}{d\rho} \left(
  \frac{y}{\rho}\right) + 2\mu
  (\varepsilon{(\mu)}-U(\rho))y=0\label{A.3}\ee
  where $U(\rho) = C_0 \sigma \rho,~~ C_0 =\frac{32\sqrt{2}
  (1+\sqrt{2})}{15\pi}$, and the total mass eigenvalue is $M= min
  (\mu+\varepsilon (\mu))$.
  Using the same approach as in \cite{28}, one can deduce the
  following relation
  \be
  |\psi(0)|^2 =\frac{\mu}{8\pi^3} \left\{ \lan
  \frac{1}{\rho^4}\ran 6 (M-\mu) - 5 \sigma C_0\lan
  \frac{1}{\rho^3}\ran \right\}\label{a.4}\ee
  where $\lan \rho^{-n}\ran = \int^\infty_0 y^2 (\rho)
  \rho^{5-n}d\rho$. From the minimization one can find the optimal
  $\mu=\mu_0 =2.23 \sqrt{\sigma_f}$ and from the  form of the
  effective potential, having minimum at $\rho=\rho_0
  =1.15/\mu_0$, one can define with good accuracy that $\lan
  \rho^{-n}\ran \approx \rho^{-n}_0$.
  Finally, taking into account the usual
relativistic normalization of bosons, $\frac{1}{\sqrt{2\mu}}$, the
total normalization of the two-gluon wave-function appears as \be
|\Psi_n(0)|^2=\frac{1}{4\mu^2_0} |\psi(0)|^2= 0.108
\sigma^2_f\label{A.5}\ee which was used in the main text.

The eigenvalue $M_0$ with the account of the hyperfine interaction
reads (see \cite{21} for details)
\be
M_0^{(2g)} = 6.15 \sqrt{\sigma_f} +\Delta M_{SS}\label{A.6}\ee
with $\Delta M_{SS} = \veS^{(1)} \veS^{(2)} 0.49 \mu_0 \frac43
\alpha_s$, and for $\veS=\veS^{(1)} +\veS^{(2)} =\mathbf{1}$, and
$\alpha_s=0.15$ one obtains the total eigenvalue, which is
$M_0^{(2g)}=6.03 \sqrt{\sigma_f} =2.56$ GeV for $\sigma_f=0.18$
GeV$^2$.

\end{document}